\documentclass{aa}  

\usepackage{graphicx}
\bibpunct{(}{)}{;}{a}{}{,}

\usepackage{txfonts}
\def\etal{\hbox{et al.}}
\def\degr{\hbox{$^\circ$}}
\def\arcmin{\hbox{$^\prime$}}

\def\sun{\hbox{$_{\odot}$}}

\def\utw{\smash{\rlap{\lower5pt\hbox{$\sim$}}}}
\def\udtw{\smash{\rlap{\lower6pt\hbox{$\approx$}}}}

\newcommand \ha{H$\alpha$ }

\newcommand \kms{km s$^{-1}$}

\newcommand \ang{\AA\,}

\newcommand \ms{M$_{\odot}$}
\newcommand \rs{R$_{\odot}$}
\newcommand \ls{L$_{\odot}$}
\newcommand \BVRI{BV(RI)$_{\textrm{\footnotesize{c}}}$ }

\begin{document}

   \title{Dynamical star-disk interaction in the young\\stellar 
          system V354 Mon\thanks{Based on the observations obtained with the 
CoRoT satellite, at the Observat\'orio Pico dos Dias, Brazil, and at the 
Observatoire de Haute Provence, France. The CoRoT space mission was developed 
and is operated by the French space agency CNES, with participation of ESA's 
RSSD and Science Programmes, Austria, Belgium, Brazil, Germany, and Spain.}}

   \author{N.N.J. Fonseca\inst{1,2,3},
           S.H.P. Alencar\inst{1},
           J. Bouvier\inst{2},
           F. Favata\inst{4},
           \and
           E. Flaccomio\inst{5}
          }

   \institute{Departamento de F\'{\i}sica -- ICEx -- UFMG, 
              Av. Ant\^onio Carlos, 6627, 31270-901,Belo Horizonte, MG, Brazil\\
              \email{nath@fisica.ufmg.br}
         \and 
              UJF-Grenoble 1 / CNRS-INSU, Institut de Plan\'etologie et 
              d'Astrophysique de Grenoble (IPAG) UMR 5274, Grenoble, F-38041, 
              France
         \and
              CAPES Foundation, Ministry of Education of Brazil, Brasília – 
              DF 70040-020, Brazil
         \and
              European Space Agency, 8-10 rue Mario Nikis, 75738 Paris 
              Cedex 15, France
         \and
              Istituto Nazionale di Astrofisica, Osservatorio Astronomico di 
              Palermo G.S. Vaiana, Piazza del Parlamento 1, 90134 Palermo, Italy
              }

   \date{Received ; accepted }

\abstract
% context heading (optional)
{}
% aims heading (mandatory)
{The main goal of this work is to characterize the mass accretion and ejection 
processes of the classical T Tauri star V354 Mon, a member of the young stellar 
cluster NGC 2264.}
% methods heading (mandatory)
{In March 2008, photometric and spectroscopic observations of 
V354 Mon were obtained simultaneously with the CoRoT satellite, the 60 cm 
telescope at the Observat\'orio Pico dos Dias (LNA - Brazil) equipped 
with a CCD camera and Johnson/Cousins \BVRI filters, and the SOPHIE 
\'echelle spectrograph at the Observatoire de Haute-Provence (CNRS - France).}
% results heading (mandatory)
{The light curve of V354 Mon shows periodical minima (P = 5.26 $\pm$ 0.50 days) 
that vary in depth and width at each rotational cycle. The \BVRI observations 
indicate that the system becomes slightly bluer as the flux increases. The 
spectra of this T Tauri star exhibit variable emission lines, with blueshifted 
and redshifted absorption components associated with a disk wind and with the
accretion process, respectively, confirming the magnetospheric accretion 
scenario. From the analysis of the photometric and spectroscopic data, it is 
possible to identify correlations between the emission line variability and the 
light-curve modulation of the young system, such as the occurrence of
pronounced redshifted absorption in the \ha line at the epoch of minimum flux. 
This is evidence that during photometric minima we see the accretion funnel 
projected onto the stellar photosphere in our line of sight, implying that the 
hot spot coincides with the light-curve minima. We applied models of cold and 
hot spots and a model of occultation by circumstellar material to investigate 
the source of the observed photometric variations.}
% conclusions heading (optional) 
{We conclude that nonuniformly distributed material in the inner part of the 
circumstellar disk is the main cause of the photometric modulation, which does 
not exclude the presence of hot and cold spots at the stellar surface. It is 
believed that the distortion in the inner part of the disk is created by the 
dynamical interaction between the stellar magnetosphere, inclined with respect 
to the rotation axis, and the circumstellar disk, as also observed in the 
classical T Tauri star AA Tau and predicted by magnetohydrodynamical numerical 
simulations.}


%\titlerunning{Analysis of star-disk interaction in the young stellar system V354 Mon}
\authorrunning{N.N.J. Fonseca \etal }
\maketitle
%
%________________________________________________________________

\section{Introduction}\label{intro}

The study of young stellar objects is important for understanding the phenomena 
that occur in star and planet formation, including our solar system. T Tauri 
stars are young ($\sim 10^{6}$ years), low-mass stars (M $\leq$ 2 M\sun) in the 
pre-main sequence (PMS) phase, which are of great interest as prototypes of 
young solar-type stars. They emit X-rays and have strong magnetic fields 
\citep[$\sim$ 2 kG,][]{johnskrull01art}. Based on their H$\alpha$ emission 
strength, they are classified as classical (CTTSs) or weak-line T Tauri stars 
(WTTSs). CTTSs have a flux excess with respect to the stellar photosphere at 
infrared, optical, and ultraviolet wavelengths, which is not observed in WTTSs.

The CTTSs also exhibit irregular photometric and  spectroscopic 
variability, with broad emission lines, strong H$\alpha$ emission, and forbidden
emission lines. Most of these general features are reproduced by magnetospheric 
accretion models \citep{shu94,hartmann94,muzerolle01,kurosawa06,lima10}, in 
which a young magnetized star accretes material from a circumstellar disk. The 
stellar magnetic field is strong enough to disrupt the disk at a distance of few
stellar radii from the star. Ionized material from the inner disk is then 
channeled onto the stellar surface along field lines, creating accretion 
funnels. Hot spots are produced at the stellar surface by the strong shock of 
material at free-fall velocity. Material is also ejected from the system in
the form of a stellar wind and as a disk wind along open magnetic field lines. 
From this model, we can explain some characteristics observed in CTTSs. Hot 
spots are responsible for optical and ultraviolet excess emission. The broad 
emission lines, which usually present redshifted absorptions, are predominantly 
produced by accelerated material in the accretion funnels. Blueshifted 
absorptions and forbidden emission lines are formed in the low-density wind. 
Infrared excess emission comes from the reprocessing by the disk of radiation 
generated in the system.

Although the magnetospheric accretion scenario describes the general CTTSs 
properties, some observational results indicate that this axisymmetric, stable 
model is not completely correct \citep{bouvier07b}. Some studies demonstrated 
that the outflow and inflow processes are intimately connected 
\citep{cabrit90,hartigan95}. The interaction between the stellar magnetosphere 
and the disk is expected to be very dynamic, as shown by magnetohydrodynamical 
(MHD) simulations \citep{goodson99,romanova02}. As a result of differential 
rotation between the star and the inner disk region where the accretion flux 
originates, the field lines may be distorted after a few rotational periods, 
eventually reconnecting and restoring the initial field configuration. This 
process repeats as the star rotates. Furthermore, a misalignment between the 
rotation and magnetic axes creates a deformation in the inner disk, leading to 
the formation of non-axisymmetric regions where accretion is favored 
\citep{terquem00,romanova03}. 

One of the best-studied CTTSs is AA Tau, observed for a month during three 
different campaigns \citep{bouvier99,bouvier03,bouvier07}. This star exhibits a 
peculiar photometric behavior, with almost constant brightness interrupted by 
quasi-cyclical and irregular episodes of attenuation. The brightness decrease 
occurs in absence of significant color variation and with an increase of 
polarization \citep{menard03}. This was interpreted as occultation of the 
stellar photosphere by circumstellar dusty material present in an inner disk 
warp. The warp is produced by the interaction between the disk and the stellar 
magnetic field, misaligned with respect to the rotation axis, as suggested by 
MHD simulations. Simultaneous high-resolution spectroscopy has shown signs of 
correlation between mass accretion and ejection processes. During the second 
observing campaign, the photometric and spectroscopic variability was 
drastically reduced for a few days, revealing an episode of disruption of 
magnetic configuration at the inner edge of the disk, which suppressed accretion
temporarily. The observed variations in H$\alpha$ absorption components 
showed the cyclical process of inflation and reconnection of field lines 
caused by the differential rotation between the star and the inner disk region, 
once again in agreement with MHD simulations. The same evidence of this 
dynamical interaction was observed again in the third campaign, five years 
after the second one.

Even though the photometric behavior of AA Tau was atypical at that time, the 
characteristics of the structure of its accretion zone might be similar in 
other CTTSs. To confirm this hypothesis, it would be necessary to obtain 
high-precision photometry for many CTTSs during many rotational periods.
This corresponds to a continuous observation of at least a month, based on 
typical rotational periods of eight days for CTTSs in Taurus. Such a monitoring 
from the ground is complicated because of telescope-time allocation and weather 
conditions. The CoRoT satellite additional program to observe the star formation
region NGC 2264 during 23 days uninterruptedly in 2008 allowed this analysis to 
be performed. 

NGC 2264 is a well-studied young stellar cluster, located in the Mon OB1 
association at $\sim$ 760 pc \citep{sung97,gillen14}, with evidence of 
active star formation \citep{dahm08}. The pioneering work of \citet{herbig54}, 
\citet{walker56}, and others established this region as an important laboratory 
for studies of star formation and  evolution of young stars. From the light 
curves of 83 CTTSs identified among the 301 cluster members observed with CoRoT 
in 2008, 23 have been classified as exhibiting the same type of variability as 
AA Tau \citep{alencar10}. This revealed that the photometric behavior of AA Tau 
is common in young stellar objects; it was found in 28\% $\pm$ 6\% of the CTTSs 
in NGC 2264 observed with CoRoT. These systems are essential to test predictions
of MHD simulations.

The main goal of the work presented in this paper is to characterize the mass
accretion and ejection properties of the CTTS V354 Mon, a member of NGC 2264,
from simultaneous high-resolution photometric and spectroscopic observations. 
V354 Mon is one of the 23 cluster members whose light curves have been 
classified as AA Tau type. From the analysis of these data, we identify 
correlations between emission line variability and light-curve modulation. It is
possible to investigate the dynamical processes of mass accretion and outflow 
that occur in this system, as well as the interaction between the stellar 
magnetic field and the circumstellar disk. We also test predictions of 
magnetospheric accretion models and MHD simulations, constructing plausible 
scenarios for the phenomena observed in this young object. 

The paper is organized as follows: Sect. 2 describes the observations and
data reduction. In Sect. 3 we present the data analysis and results. In 
Sect. 4 we discuss the possible causes of the observed photometric and 
spectroscopic variations. The conclusions are depicted in Sect. 5.

%__________________________________________________________________

\section{Observations}\label{obs}

CoRoT observed NGC 2264 from March 7 to 30, 2008. For V354 Mon, a cluster member
in the CoRoT observational program, high-resolution spectroscopy and \BVRI 
photometry were obtained simultaneously with the SOPHIE \'echelle spectrograph 
at the Observatoire de Haute-Provence (OHP - CNRS, France) and the 60 cm 
telescope equipped with a CCD camera and Johnson/Cousins filters at the 
Observat\'orio Pico dos Dias (OPD - LNA, Brazil), respectively. The journal of 
observations is given in Table \ref{t:regobs}. V354 Mon is a CTTS with a 
well-determined photometric period, which increases the chances that this system
is being seen edge-on, and exhibits a large \ha equivalent width, which 
facilitates the study of its spectroscopic variability with good signal-to-noise
ratio, at least in this line. Information from the literature about V354 Mon is 
gathered in Table \ref{t:infov354}.

\begin{table}[ht]
\caption{Journal of observations.}
\label{t:regobs}
\centering
\begin{tabular}{lccc}
\hline\hline
Dates (March 2008) & Instrument & Exp. time (s) & N$_{obs}$ \\
\hline
{\bf Spectroscopy}                  &              &        &      \\
12, 14-1, 14-2,                     & SOPHIE       & 3600   &    10\\
18-1, 18-2, 20,                     & (OHP)        &        &      \\
23, 25, 27, and 28                  &              &        &      \\
\hline
{\bf Photometry}                    &              &        &      \\
 7 to 30                            & CoRoT        & 512    & cont.\\
19, 20, 25, 26, and 27              & 60 cm        & B  400 &    10\\
                                    & (OPD)        & V  300 &     8\\
                                    &              & R  200 &     8\\
                                    &              & I  200 &    10\\
\hline
\end{tabular}
\tablefoot{On March 14 and 18, we obtained two spectra of V354 Mon, referred to 
as 14-1, 14-2 and 18-1, 18-2.}
\end{table}

Many results presented in this work are dependent on the effective temperature 
of the star. \citet{lamm05} assigned a K4 spectral type to V354 Mon, which 
agrees with the $T_{\rm eff}$ value of 4590 K obtained by \citet{flaccomio06}, 
using the \citet{kenyon95} spectral type/intrinsic color scale that is, however,
more appropriate for main-sequence dwarfs than young stars. V354 Mon was 
observed in 2011 with the FLAMES spectrograph (VLT/ESO) during a multiwavelength
campaign of the NGC 2264 star-forming region \citep{cody14} when 22 
medium-resolution spectra (R = 17\,000) were acquired. To check the V354 Mon 
spectral type, we estimated the effective temperature of V354 Mon with the 
FLAMES spectra, using line ratios. A series of synthetic spectra were calculated
with the code Spectroscopy Made Easy (SME) \citep{valenti96}, using the same 
resolution as the FLAMES observations, within the spectral domain of the FLAMES 
data (6440 \ang $< \lambda <$ 6820 \ang) and the 3500 K to 6000 K temperature 
range, with $\log{g}=4.0$ and 4.5 and solar metallicity. In this range of 
$T_{\rm eff}$ and $\log{g}$, FeI lines are very $T_{\rm eff}$ dependent, while CaI 
lines are not. The CaI/FeI ratio is therefore a good $T_{\rm eff}$ indicator that 
is also independent of veiling. We analyzed the CaI 6717.7 \ang and FeI 6546.2 
\ang lines that were always present and easily identified in the spectra. 
Comparing the line ratios measured in the observed and theoretical spectra, we 
obtained for V354 Mon $T_{\rm eff}=4647 \pm 161$ K with $\log{g}=4.5$ (dwarfs) 
and $T_{\rm eff}=4434 \pm 133$ K with $\log{g}=4.0$ (young stars). The errors come
from the standard deviation of the values obtained with all the FLAMES spectra. 
Recently, \citet{pecaut13} compiled tables of effective temperature vs. spectral
type for both dwarfs and young (5-30 Myr) stars. We compared the $T_{\rm eff}$ 
values we obtained for V354 Mon with their respective tables, and they are both 
consistent with a K4 spectral type. Recently, \citet{marinas13} assigned a K7 
spectral type to V354 Mon (their target 26) based on an [OH]/[MgI] line ratio 
calibration obtained from {\it H} band low-resolution spectra. They acknowledged
an error of two subclasses to spectral types obtained with their calibration. 
However, in the K7-K3 spectral range, the OH (1.69 $\mu$m) line is about 5 to 10
times shallower than the MgI (1.50 $\mu$m) line, which can complicate its 
evaluation in low-resolution spectra. This, together with the scatter present in
their calibration, possibly adds some more uncertainty to their spectral type 
determination in the K7-K3 region. Our results based on the FLAMES spectra are 
not compatible with a K7 spectral type, which would require much lower 
$T_{\rm eff}$ values \cite[4050 K for dwarfs and 3970 K for young stars, according
to the tables of][]{pecaut13} than we obtained. We decided therefore to adopt a 
K4 spectral type for V354 Mon, in agreement with the results of \citet{lamm05} 
and \citet{flaccomio06}.

\begin{table}[h]
\caption{Characteristics of V354 Mon from the literature.}
\label{t:infov354}
\centering
\begin{tabular}{lc}
\hline\hline
Data & Ref.\\
\hline
V = 14.45 mag & 1\\
spectral type: K4V & 2\\
photometric period = $5.22 \pm 0.87$ days & 2\\
\ha equivalent width = $16.60$ \ang & 1\\
\ha width at $10\% > 270$ \kms & 3\\
log(T$_{\textrm{eff}}$/K) = $3.66$ & 4\\
log(L$_{\textrm{bol}}$/\ls) = $0.34$ & 4\\
mass = $1.50$ \ms & 4\\
log(age/years) = $6.42$ & 4\\
heliocentric radial velocity (Dec 2004): $19.42$ \kms & 3\\
\hline
\end{tabular}
\tablebib{(1) \citet{dahm05};
(2) \citet{lamm05}; (3) \citet{furesz06}; (4) \citet{flaccomio06}.
}
\end{table}

\subsection{Photometry}

The CoRoT data were reduced using a standard procedure 
\citep{samadi07,auvergne09} and delivered in the form of a light curve. The 
light curve was processed using a sigma-clipping filter that removed outliers 
(mostly related to South Atlantic Anomaly crossings) and hot pixels, and 
corrected for the effects associated with entrance into and exit from Earth 
eclipses. The light curve was also rebinned to 512 s and corresponds to the 
integrated flux in the CoRoT mask.

The OPD photometric observations were carried out on five nights between March 
19 and 27, 2008. The 60 cm telescope was equipped with the CCD camera \#106 
(SITe SI003AB) of $1024\times1024$ pixels and Johnson/Cousins \BVRI filters. 
Integration times ranged from 200 s to 400 s, depending on the filter. We
processed raw images with the usual techniques within the IRAF environment, 
which included bias subtraction and flat-field calibration. Differential 
photometry between V354 Mon and a reference star was also performed using IRAF. 
The reference star is TYC 750-1637-1 (spectral type F7V, V = 11.609; 
\citealt{sung97}), located less than 2.2\arcmin\, away from V354 Mon and 
recorded on the same images.

\subsection{Spectroscopy}\label{obs_spec}

The spectroscopic observations were collected on eight nights between March 12 
and 28, 2008. We obtained ten high-resolution spectra at the 1.93 m OHP 
telescope with the SOPHIE dual fiber échelle spectrograph \citep{perruchot08}, 
which yields 41 orders covering the 3872 \ang to 6943 \ang domain at a mean 
spectral resolution of $\lambda / \Delta \lambda \simeq 75\,000$ 
(high-resolution mode), and simultaneously records the object and the 
neighboring sky. All spectra have the same exposure time of 3600 s. The data 
were automatically reduced via a standard procedure, adapted from the software 
used with the HARPS spectrograph designed at the Geneva Observatory. The 
reduction procedure includes bias subtraction, optimum extraction of the orders,
removal of cosmic rays, flat-fielding achieved through a tungsten lamp exposure,
wavelength calibration with a thorium lamp exposure, cross-correlation with an 
appropriate numerical mask, and reconnection of spectral orders with barycentric
correction. 

Unfortunately, the nebular emission present in the stellar spectrum and in the 
sky spectrum are not equivalent, because the fibers used in the observations do 
not respond equally to the same light source\footnote{This problem was first
investigated by selecting only nebular emission lines and calculating the ratio 
between the highest intensity of these lines in the sky and stellar spectra. It 
was later confirmed by OHP technicians through private communication.}. In 
addition, nebular emission is highly variable on scales of arcseconds or less, 
according to studies of HII regions \citep{henney99,mccollum04}. For these 
reasons, contamination by nebular emission was hard to exclude properly and some
emission lines could not be analyzed. For each spectrum, we subtracted the sky 
continuum level, normalized the stellar continuum to unity, and corrected from 
the star radial velocity. Then we subtracted a template spectrum (HD190007) of 
same spectral type as V354 Mon, resulting in circumstellar spectral profiles. We
also calculated the average profile of the spectra obtained on the same night, 
since there is no significant variation between them, resulting in one spectrum 
for each night of observation.

%__________________________________________________________________

\section{Results}

\subsection{Photometry}\label{res_phot}

\begin{figure}
\centering
\includegraphics[width=9cm]{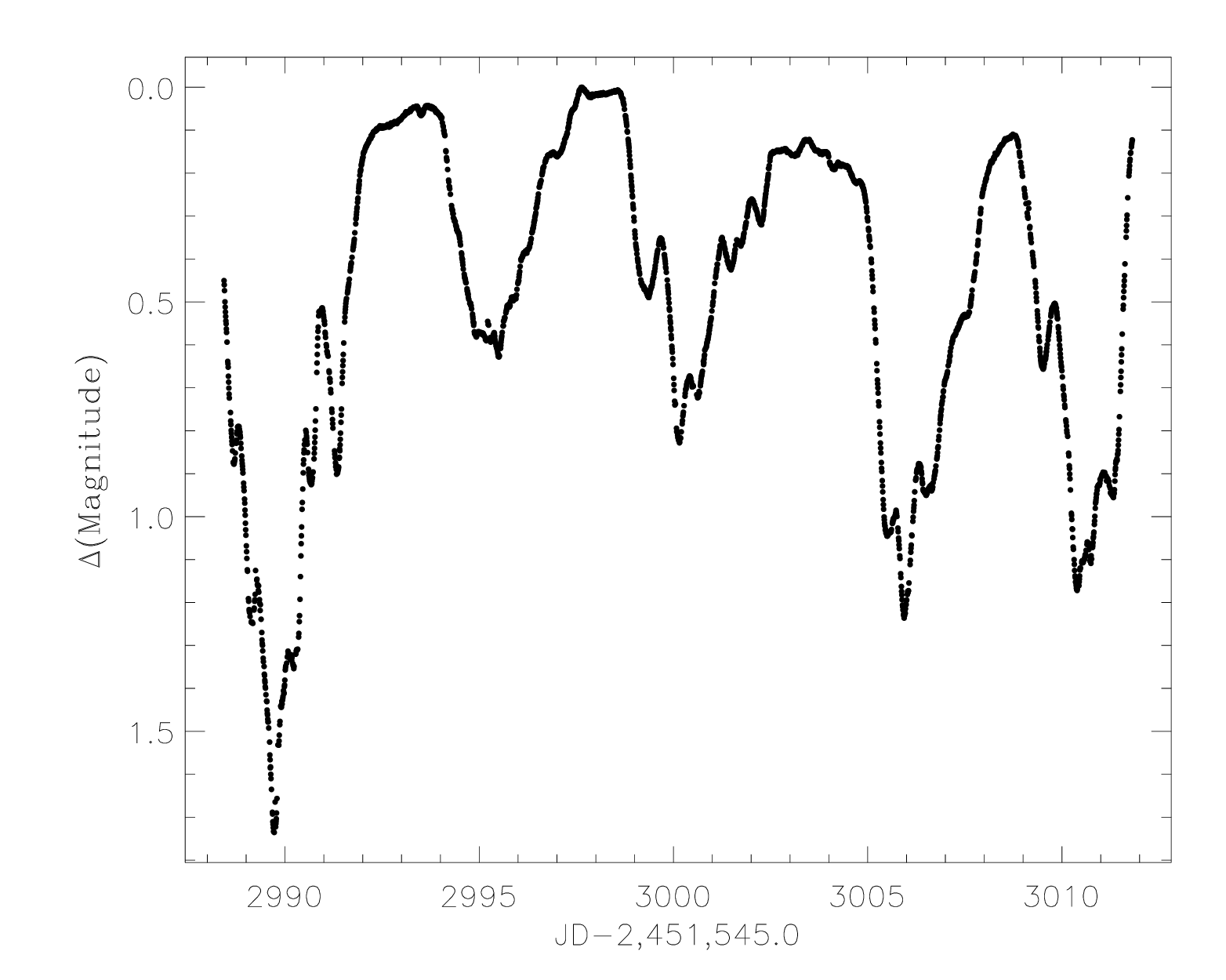}
\includegraphics[width=9cm]{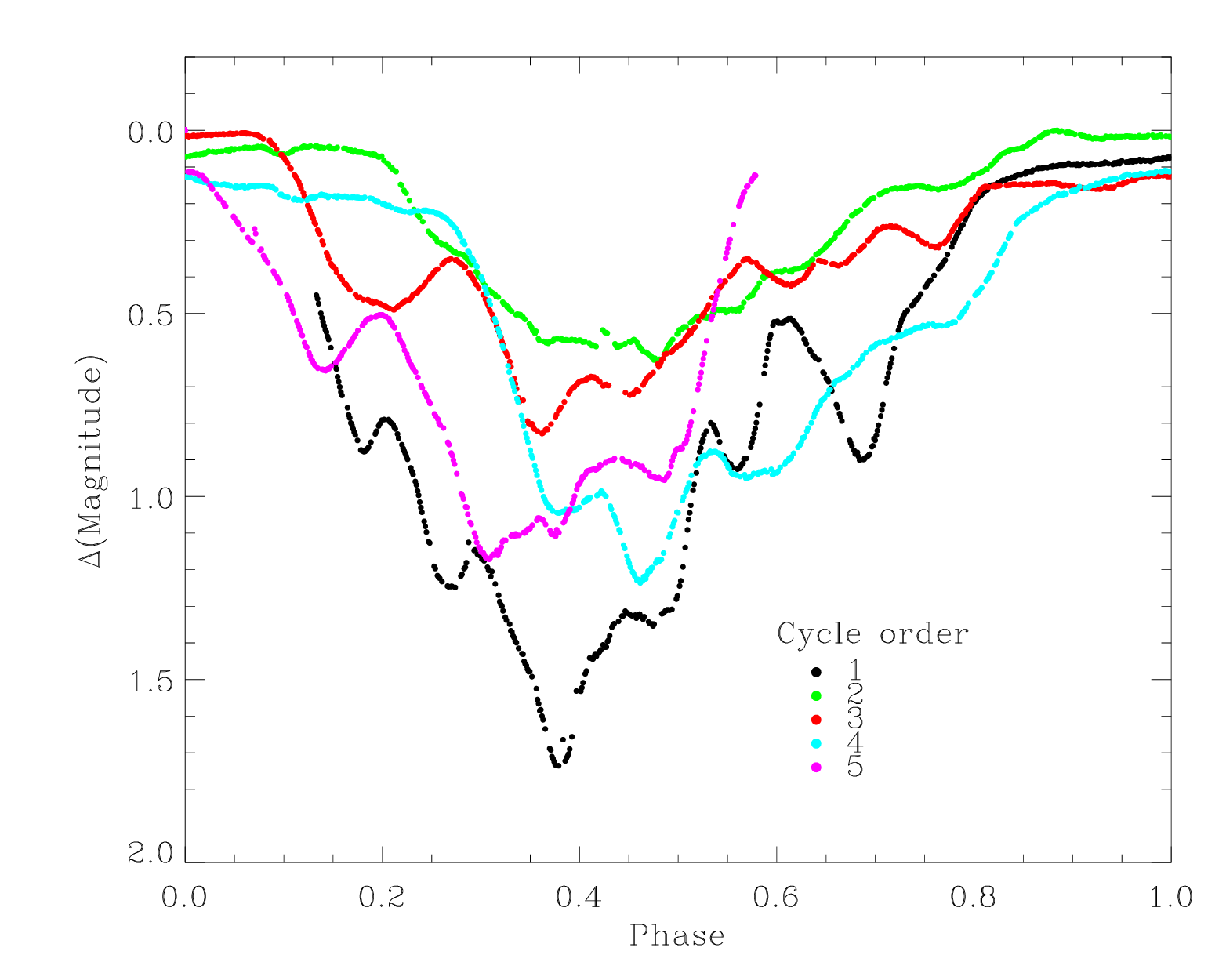}
\caption{CoRoT light curve, continuous (top) and folded in phase (bottom) 
with a period of $5.26 \pm 0.50$ days. Different colors represent different 
rotational cycles in the order indicated in the lower right corner of the bottom
panel. Magnitudes are given on an arbitrary scale.}
\label{f:corotlc}
\end{figure}

The broadband, white-light curve of V354 Mon obtained with CoRoT (Fig. 
\ref{f:corotlc}, top panel) displays a typical CTTS behavior, noticeably 
periodic and with well-defined minima and maxima that vary in depth and width 
from one rotational cycle to the other. A periodogram analysis \citep{scargle82}
of the light curve reveals a period of $5.26 \pm 0.50$ days, which is consistent
with the value obtained by \citet{lamm05}, $5.22 \pm 0.87$ days. This indicates 
that the dominant source of photometric variability did not significantly change
on a timescale of a few years. The varying depth and width at minimum is more 
evident when the light curve is folded in phase with the calculated period 
(Fig. \ref{f:corotlc}, bottom panel).  

We measured the percentage variability amplitude of the light curve as 
[(Flux$_{max}$ - Flux$_{min}$)/Flux$_{median}$] $\times$ 100, obtaining a value 
of 115\% for V354 Mon. This is one of the largest photometric variations in the 
CTTSs observed with CoRoT, which range between 3\% and 137\% \citep{alencar10}.

\begin{figure}
\centering
\includegraphics[width=9cm]{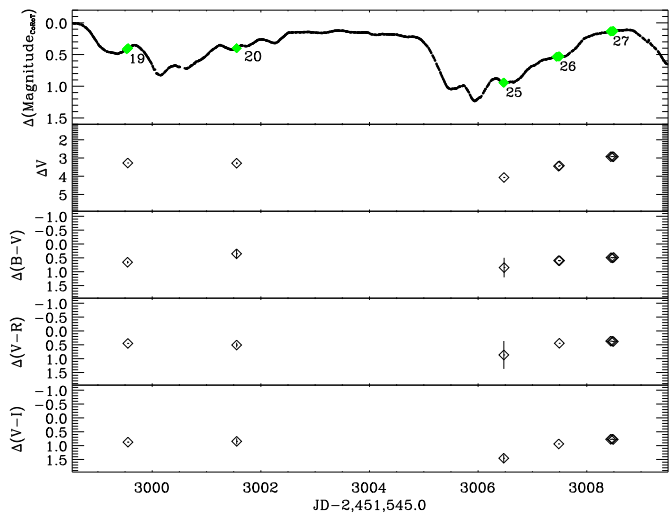}
\caption{CoRoT light curve, marked with OPD photometric observation dates, 
compared with the V-band photometry and color variations of V354 Mon. Vertical 
bars indicate the errors of each measurement. The magnitude scale on the 
vertical axis is the same for each color plot.}
\label{f:corotlna}
\end{figure}

As observed in Fig. \ref{f:corotlna}, the OPD data are consistent 
with the CoRoT photometry. From the calculation of magnitude differences 
between the observations of March 25 and 27, we note that the photometric 
amplitude decreases toward longer wavelengths: $1.5 \pm 0.3$ in B, 
$1.15 \pm 0.05$ in V, $0.6 \pm 0.5$ in R, and $0.4 \pm 0.1$ in I. Analyzing the 
color variation (Fig. \ref{f:corotlna}), we see that the system becomes slightly
bluer as the flux increases.

\subsection{Spectroscopy}

For the reasons stated in Sect. \ref{obs_spec}, contamination by nebular 
emission present in the stellar spectrum was hard to exclude properly, 
which restricted the analysis of emission lines. In this work we focus on the 
\ha region. To exclude the nebular emission from calculations or identify 
its contribution in the results, we defined the range of the contaminated 
spectral region. In each sky spectrum we determined the wavelength values 
that constrain the region around 6562.85 \ang where the flux is higher than the 
normalized background continuum. The mean values obtained are 6561.4 and 6563.4
\ang, corresponding to velocity values of $-64.6$ and $27.1$ \kms, respectively,
relative to the spectral line center at the stellar rest frame. 

\begin{figure}
\centering
\includegraphics[angle=+90,width=8cm]{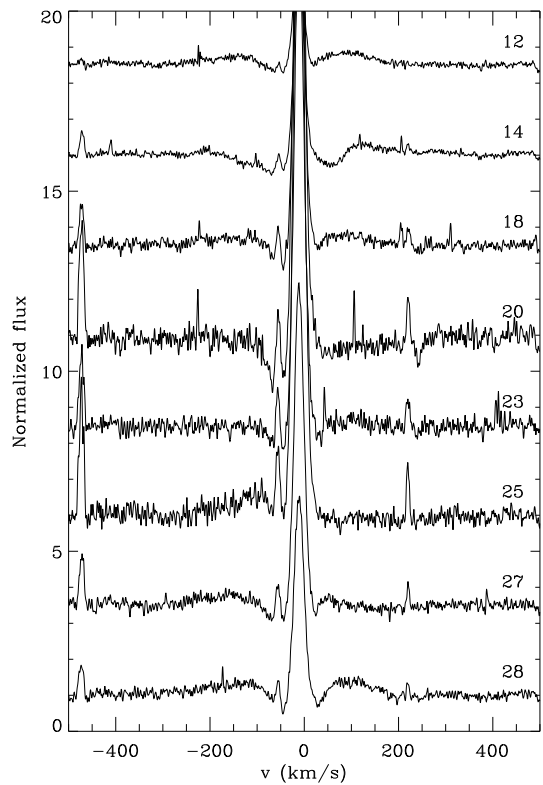}
\caption{\ha circumstellar spectra for each observed night, indicated at the
right of each profile. The continuum of each spectrum has been normalized to 
unity and the profiles have been shifted for clarity.} 
\label{f:haspec}
\end{figure}

\begin{figure}
\centering
\includegraphics[width=9cm]{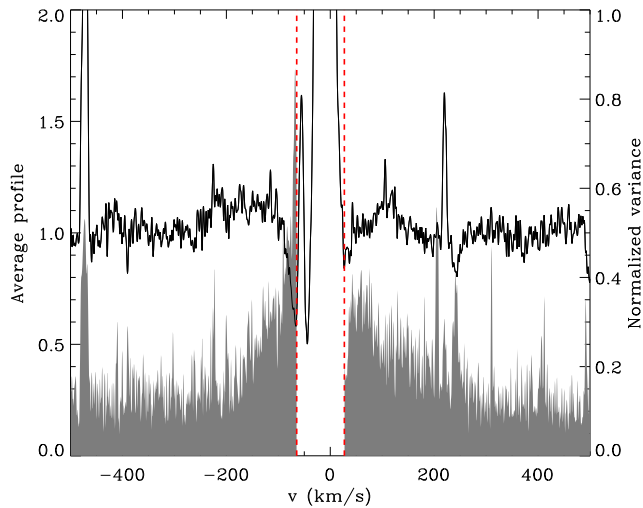}
\caption{Average profile (solid line) and normalized variance (gray shaded area)
of \ha line. Dashed red lines delimit the region dominated by nebular emission.}
\label{f:medvarha}
\end{figure}

The circumstellar profiles exhibit a remarkable variability (Fig. 
\ref{f:haspec}), with a well-distinguishable nebular emission in the central 
part of the plots. We show in Fig. \ref{f:medvarha} the \ha line average profile
with its normalized variance \citep{johns95}, which measures how much each 
region of the observed profiles varies with respect to the average profile. We 
note that the blueshifted region is more variable than the redshifted one. The
blueshifted absorption seen in the average profile is generally associated with 
winds in the system.

\begin{figure}
\centering
\includegraphics[width=9cm]{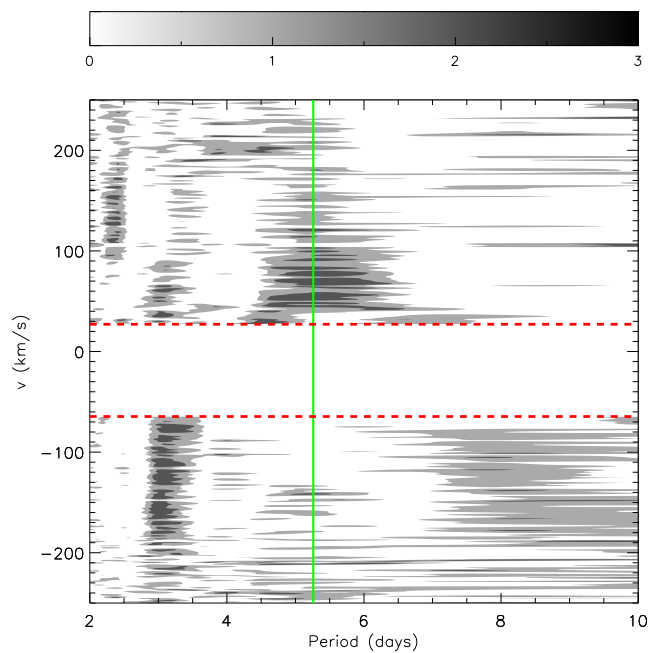}
\caption{Two-dimensional periodogram of \ha line flux. The power scale varies 
from zero (white) to the highest value (black). Dashed red lines delimit the 
region dominated by nebular emission. The solid green line marks the period of 
5.26 days.}
\label{f:hapercorr}
\end{figure}

Using the same method as was applied to calculate the light-curve period 
discussed in Sect. \ref{res_phot}, we investigated the \ha normalized flux 
periodicity through a periodogram analysis of the observed time series, which we
did independently in each velocity bin of 0.5 \kms along the profile. The 
results were disposed side by side in space velocity to form an image. The 
two-dimensional periodogram is shown in Fig. \ref{f:hapercorr}, where the 
normalized power scales from zero (white) to the highest value (black). The \ha 
redshifted side displays periodicity in a broad region centered on 5.3 days, 
which is close to the photometric period, while the blueshifted side is 
variable at around 3.1 days.

The period obtained for the variation of the \ha redshifted side is close to 
the photometric period, which is interesting from the point of view of 
magnetospheric accretion with a misalignment between the rotation and magnetic
axes. In this scenario, material from the inner part of the disk is more easily 
channeled along the field lines that are deflected from the disk mid-plane
towards the stellar surface because of the tilt of the stellar magnetosphere. 
Thus there is a preferential region of accretion, causing the \ha redshifted 
absorption to vary with the stellar rotational period.  This absorption will be 
more pronounced and at higher velocities when we see the accretion region 
projected onto the stellar photosphere along our line of sight.

The modulation of the \ha blueshifted side presents a period shorter than the 
photometric period. In this case, the existence of only one preferential region
of ejection of material from the disk is not physically acceptable, as the wind 
generation is located immediately beyond the accretion region, at a slightly 
larger distance from the star, which corresponds to a Keplerian period also 
slightly longer. The period of $\sim$ 3 days may indicate that in fact we 
observe two major contributions of the wind, at opposite sides. Consequently, 
the formation region of these components is related to a Keplerian period of 
$\sim$ 6 days, slightly longer than the period of the variation of \ha 
redshifted side, which is connected with the accretion process.

From analyzing the \ha profiles, we observe that the spectrum obtained on March 
25 shows a strong emission in the blueshifted side and no emission in the 
redshifted side. Because no other profile resembles it, this blueshifted 
emission might be a sign of a larger, occasional ejection of material, maybe 
related to the disruption of the magnetic field configuration. Since this 
isolated event is not linked to the rotational modulation of the system, we made
a new analysis of the \ha flux periodicity excluding the spectrum of March 25. 
We recovered the results obtained considering all spectra, but the blueshifted 
side also presented a broad region around the photometric period, while a 
periodicity of about 3 days in the redshifted side becomes evident, making the 
new 2D periodogram symmetric. Then all \ha components shows periodic variation
according to the photometric modulation. However, the period of 3 days has to be
confirmed with a richer dataset because we have only a few points to investigate
additional periods.

\begin{figure}
\centering
\includegraphics[width=8.5cm]{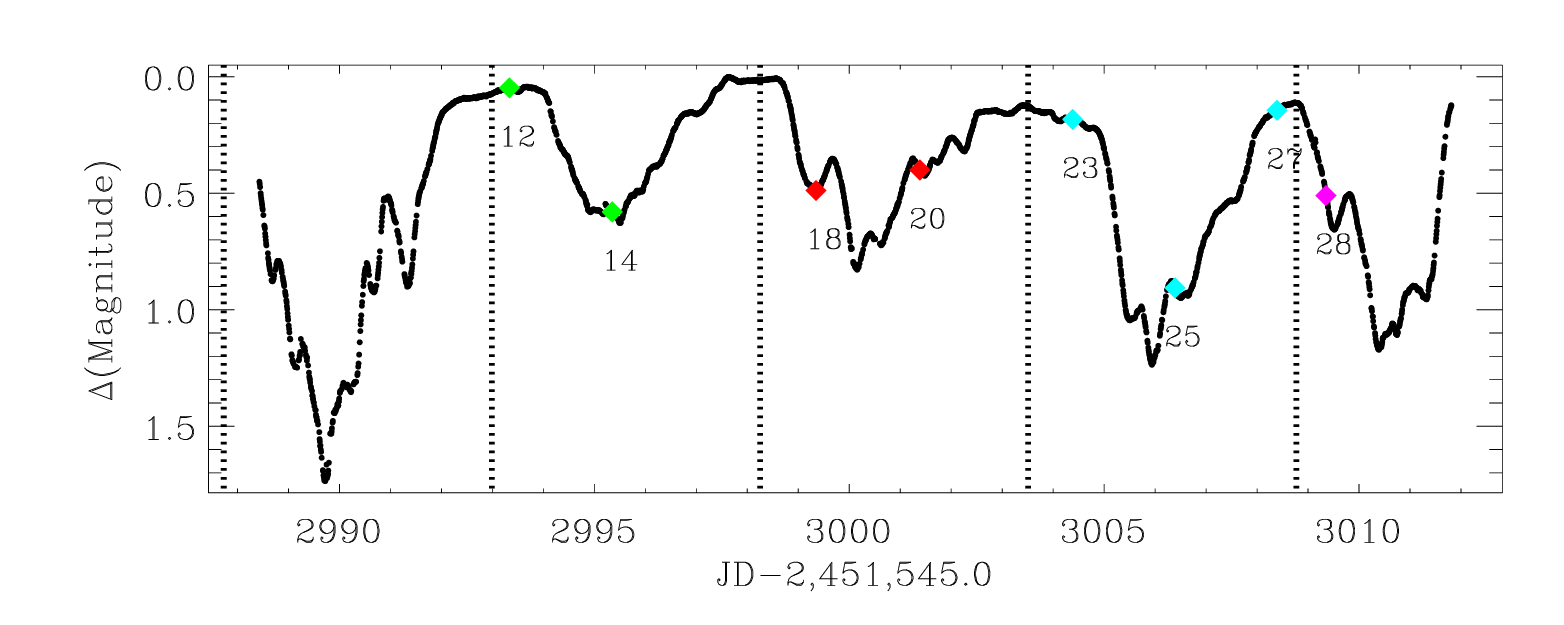}
\includegraphics[angle=+90,width=7.5cm]{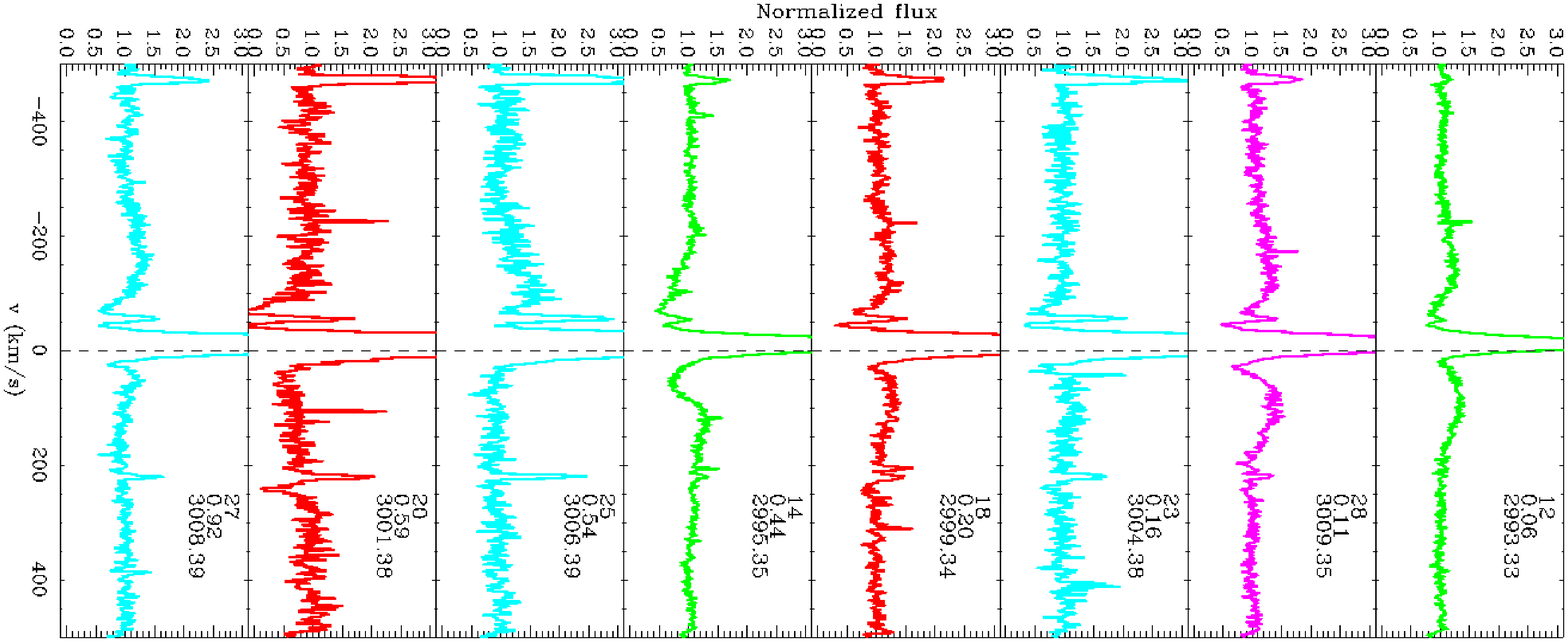}
\caption{\ha profiles ordered according to rotational phase (middle number in 
each panel). The observation date and CoRoT JD are also displayed (top and
bottom numbers). The color code is the same as in Fig. \ref{f:corotlc}, bottom 
panel. The dashed vertical line marks the central position of the line, 
corresponding to 6562.85 \ang. The top panel helps to locate the spectroscopic 
observations in the light curve and to identify the correlation between them.}
\label{f:hanadfase}
\end{figure}

The \ha profiles are shown as a function of rotational phase with photometric 
period (Fig. \ref{f:hanadfase}) to identify similarities and differences between
the spectra that are close in phase position. The top plot helps to locate the 
spectroscopic observations in the light curve and to identify a correlation 
between them. The profile of March 28 resembles the profile of March 18, with a 
slight difference in the redshifted absorption component. They are close to each
other in phase, 0.11 and 0.20, and at similar points of the light-curve profile,
in which there is a flux decrease before a small local peak. Similarly, the 
spectra of days 12, 28, 23, and 18 are similar, with an asymmetric profile and a
more intense redshifted side. All of them are located in the first half of the
photometric cycle, in which the stellar brightness is reduced. On the other 
hand, the spectra of days 25, 20, and 27 are also similar, with an asymmetric 
profile as well, but are more intense in the blueshifted side. These 
observations are located in the flux increase of the light curve. The spectrum 
of March 14 seems to be a transition between these two situations, because it is
situated in a photometric minimum and presents a distinct profile. We conclude 
that there is a correlation between the light-curve modulation and the spectral 
line variability. The phenomenon that produces these variations appears to be 
asymmetric, since the brightness increase in the light curve seems to be slower 
than the decrease and the emission profiles observed in these two phase ranges 
exhibit different characteristics.

%__________________________________________________________________

\section{Discussion}

The most often discussed causes of photometric variations at optical 
wavelengths in young stars are cold spots produced by magnetic activity, hot 
spots created by the shock of the accretion flow at the stellar photosphere, and
partial occultation of the star by inhomogeneous circumstellar material 
\citep{herbst94}. Of these, only cold spots are not associated with an accretion
disk. We analyze below the photometric and spectroscopic variability expected in
each one of these scenarios and compare this with the observed characteristics 
of V354 Mon in an attempt to identify the physical phenomenon that dominates the
observed optical variability in this system.

\subsection{Cold and hot spots}

Cold spots are one of the most common sources of photometric variability in 
low-mass stars at all ages. They are good indicators of magnetic activity, since
they are associated with the eruption of magnetic flux from the stellar interior
out into the atmosphere. Hot spots in young stars are caused by infalling gas at
the stellar photosphere, which is a direct consequence of accretion.

Spots rotate with the star, as they are located in the photosphere, and generate
periodic photometric variability on timescales of the stellar rotational period.
If a cold spot is the main cause of light-curve modulation, it is fully visible 
to the observer at light-curve minimum, while a hot spot is fully visible at 
maximum. Both spots produce a modulation in the photometric amplitude that 
increases toward shorter wavelenghts because of the temperature difference 
between the spot and the photosphere, which makes the star bluer as it becomes 
brighter. But this effect is more pronounced for hot than for cold spots. In 
this manner, we can estimate the spot temperature and size by comparing the 
total photometric amplitude of variation at different wavelengths.

We applied the model developed by \citet{bouvier93}, which derives the 
temperature and the smallest size of a spot that is responsible for the 
modulation of the stellar brightness through the fit of amplitude variability as
a function of wavelength. This model searches for the best spot configuration 
that reproduces the observed amplitudes using a $\chi^2$ method, taking into 
account limb-darkening effects and simultaneously fitting the amplitudes in all 
bands. No assumption is made about the spot number and shape, but it is assumed 
that all have the same temperature and the temperature distribution in each 
individual spot is uniform. Therefore this model does not determine the spot 
location over the star and provides only a lower limit for the fractional area 
of the visible stellar hemisphere covered by spots.

We considered that the star has an effective temperature of 4500 K 
\citep{flaccomio06}, with limb-darkening coefficients consistent with this 
temperature given by \citet{claret00} for $\log g \sim 4.0$ and $\log$[M/H] 
$\sim$ 0. Evaluating spots with temperatures of 
$3.0 \leq \log(T_s/\textrm{K}) \leq 4.0$ and fractional area $f$ of the visible 
stellar hemisphere between 0.1\% and 90\%, the model converged to a spot of 
$T_s=10\,000$ K and $f=5\%$ as the best configuration that reproduces the 
observed amplitudes of V354 Mon. Even though this solution is at the boundary of
model parameters because it presents the highest possible spot temperature, this
result is plausible because it agrees with hot spot parameters derived for
other T Tauri stars \citep{bouvier95}.

Although the spot model indicates that hot spots can reproduce the amplitudes of
variation better than cold spots, we restricted the fit to spots with 
$T_s < 4500$ K to confirm that cold spots cannot be considered as the main cause
of photometric modulation. In this case, we obtained $T_s=3981$ K and $f=89\%$ 
as the best configuration. But this result is implausible because cold spots in 
TTSs typically cover less than $50\%$ of the projected stellar disk 
\citep{bouvier95}. We conclude that cold spots probably exist on the stellar 
surface of V354 Mon, but are not the main cause of the light-curve variability.

The effective temperature of V354 Mon adopted from \citet{flaccomio06} was 
derived based on the \citet{kenyon95} spectral type/intrinsic color scale, which
is more adequate to main-sequence dwarfs than young stars. As discussed in 
Sect. \ref{obs}, \citet{pecaut13} have recently compiled a table of effective 
temperature vs. spectral type for 5-30 Myr old PMS stars. According to this 
scale, a K4 star presents an effective temperature of 4330 K. Considering this 
value, the spot model yields results that are very similar to those obtained for
4500 K: $T_s=10\,000$ K and $f=4\%$ as the best configuration that reproduces 
the observed amplitudes of V354 Mon, and $T_s=3852$ K and $f=89\%$ if we 
restrict the model to cold spots only.

\begin{figure}
\centering
\includegraphics[width=9cm]{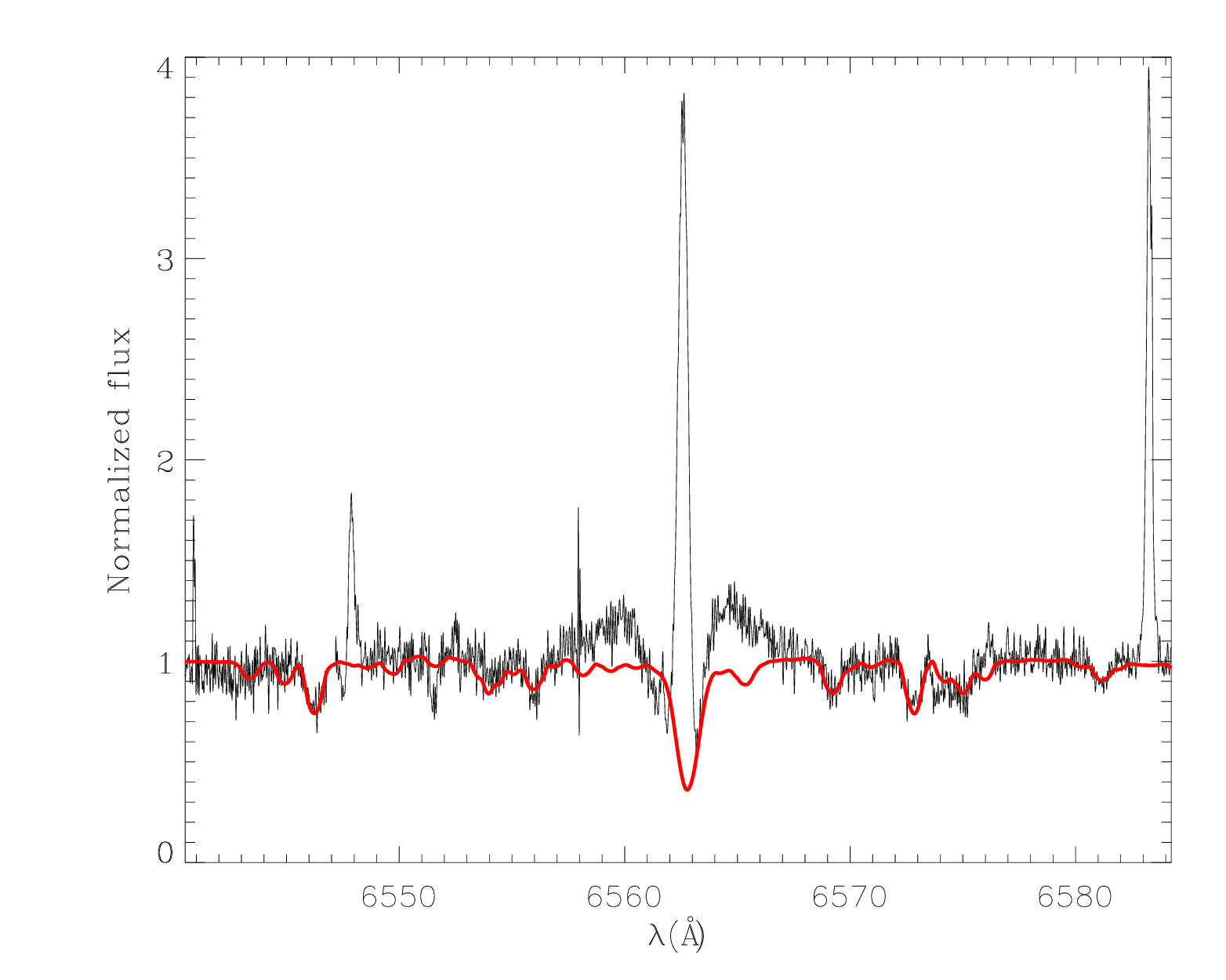}
\includegraphics[width=9cm]{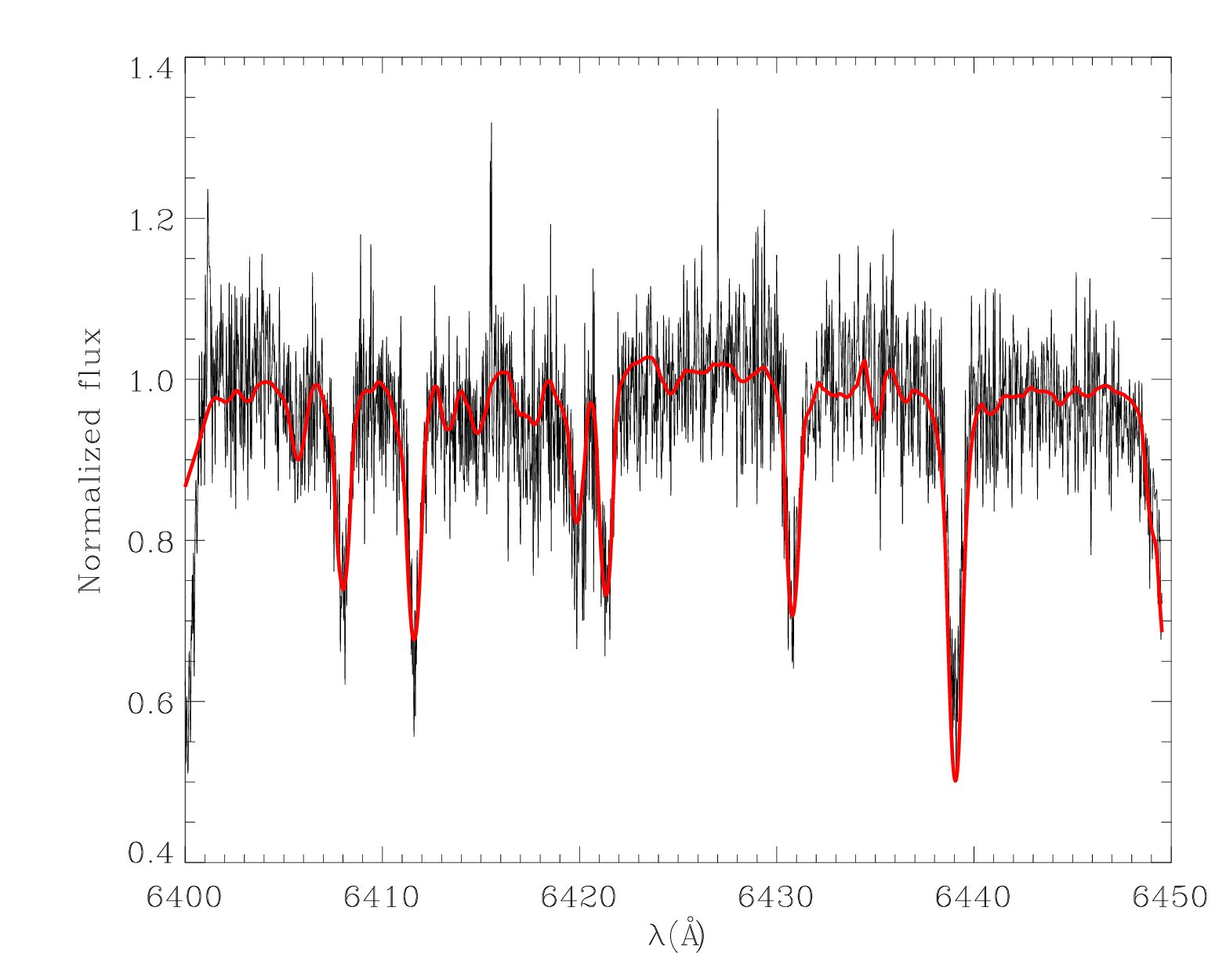}
\caption
{Two regions of the V354 Mon spectrum (black line) obtained at a photometric 
maximum on March 12, superimposed on the standard star spectrum (red line), 
corrected for its radial velocity and rotationally broadened to the V354 Mon 
value. The photospheric absorptions in the spectrum of both stars agree well, 
indicating that there is no measurable veiling in the spectrum of V354 Mon. In 
the top plot, which represents the \ha region, we observe the nebular 
contribution to this line in the V354 Mon spectrum and the photospheric 
contribution to \ha in the standard spectrum.}
\label{f:compad}
\end{figure}

\begin{figure}
\centering
\includegraphics[width=9cm]{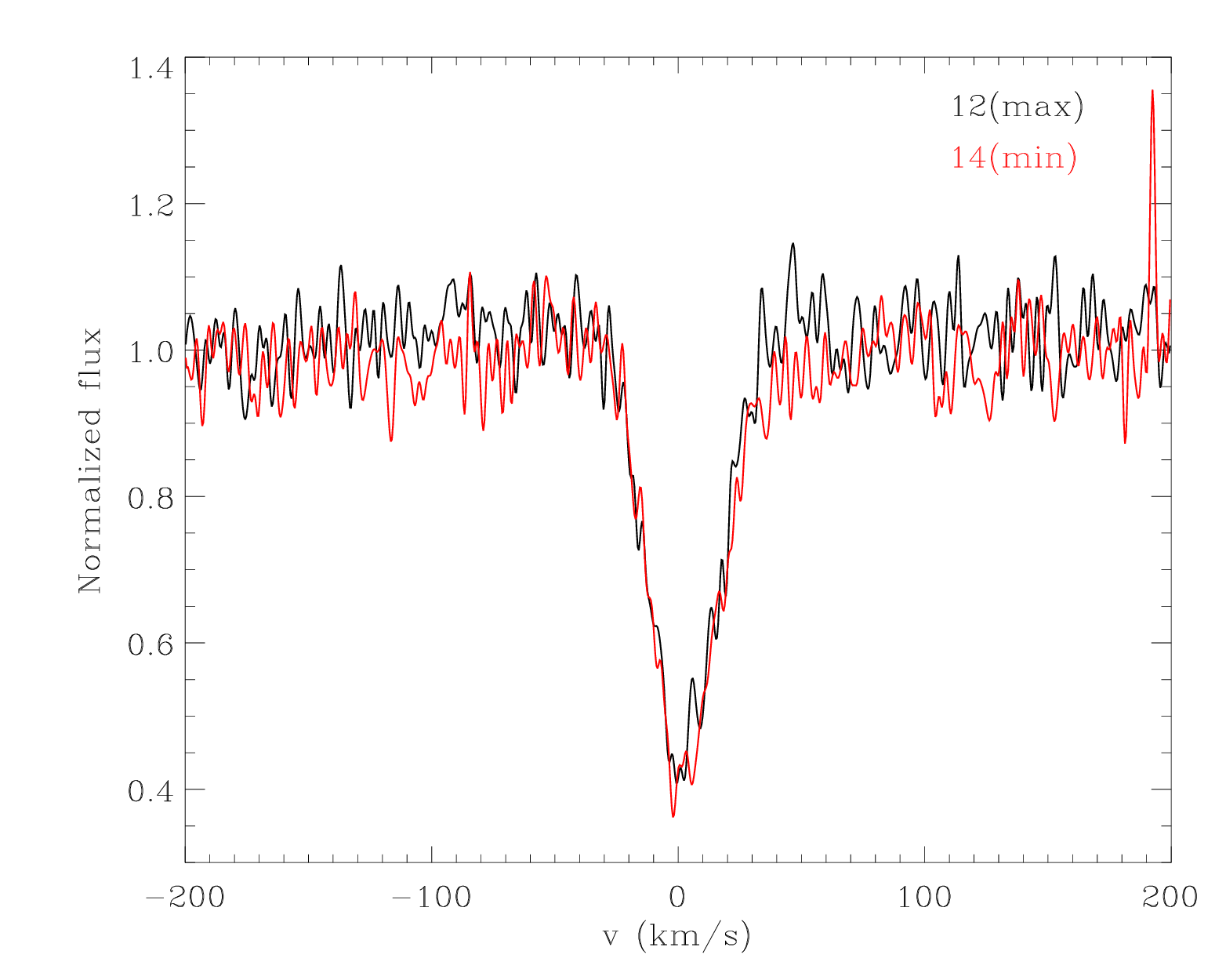}
\includegraphics[width=9cm]{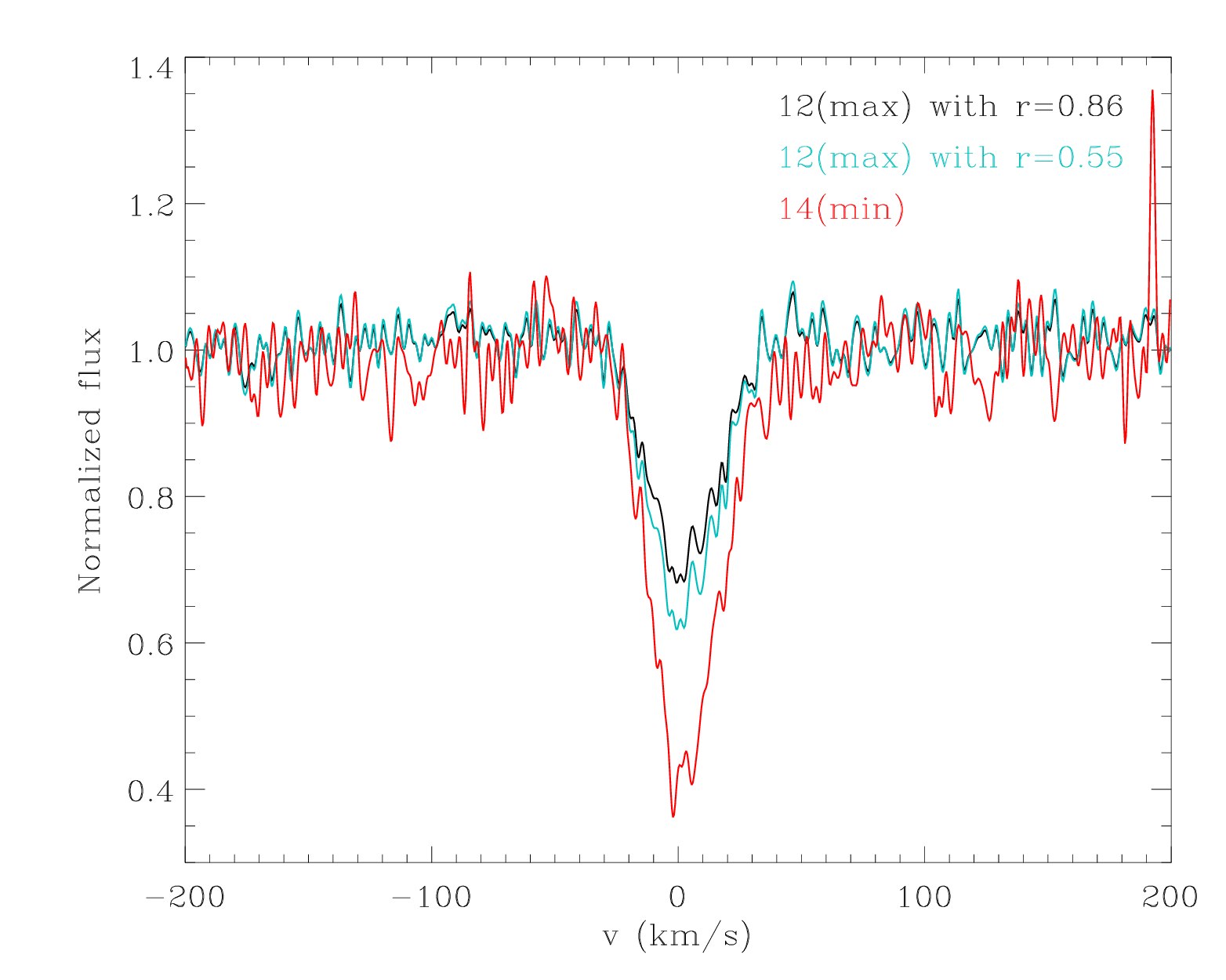}
\caption
{Comparison between LiI absorption lines observed on March 12 and 14, 
corresponding to photometric maximum and minimum. In the top plot there is no 
difference in line depth, showing that no veiling is detectable with brightness 
increase. In the bottom plot we added veiling to the spectrum at photometric 
maximum to demonstrate how a spot with temperature of $10\,000$ K and $f=5\%$ 
would affect the line depth if it were the main cause of the observed 
photometric variations.}
\label{f:liveil}
\end{figure}

A hot spot influences not only the light curve of the star, but also its 
spectrum because the accretion shock produces a continuum flux excess with 
respect to the stellar photosphere that veils the observed photospheric lines, 
reducing their depth. During the process of data reduction, we found no 
evidence of veiling, as shown in Fig. \ref{f:compad}. We calculated the flux
generated by a hot spot of $10\,000$ K and $f=5\%$, the best spot 
configuration that reproduces the observed amplitudes of V354 Mon in the \BVRI 
bands, through a blackbody curve integration for $10\,000$ K convolved with 
passband response functions \citep{bessell83}. The ratio between the values 
obtained and the stellar blackbody flux without a spot at 4500 K was 2.84 in
B, 1.34 in V, 0.86 in R, and 0.55 in I. This means that this hot spot should 
produce a remarkable veiling that would be measurable when comparing absorption 
lines observed at photometric maximum and minimum. We note in the top panel of 
Fig. \ref{f:liveil} that the LiI line, one of the best-defined absorptions in 
the spectrum, displays no difference in depth between these two occasions. To 
exemplify that the flux of a spot of $10\,000$ K and $f=5\%$ would cause a 
visible effect, we added the calculated veiling to the LiI absorption observed 
at photometric maximum, when the hot spot would be fully visible. We applied the
values obtained for R and I bands, as they have characteristic wavelengths of 
5925 \ang and 7900 \ang, and the LiI line is located at 6707.8 \ang. The result 
is shown in the bottom panel of Fig. \ref{f:liveil}, in which we observe that 
there is a noticeable decrease in the line depth due to the added veiling. The 
fact that we do not observe this in the spectra of V354 Mon refutes a hot spot 
as the main cause of photometric variability.

\begin{figure}
\centering
\includegraphics[width=9cm]{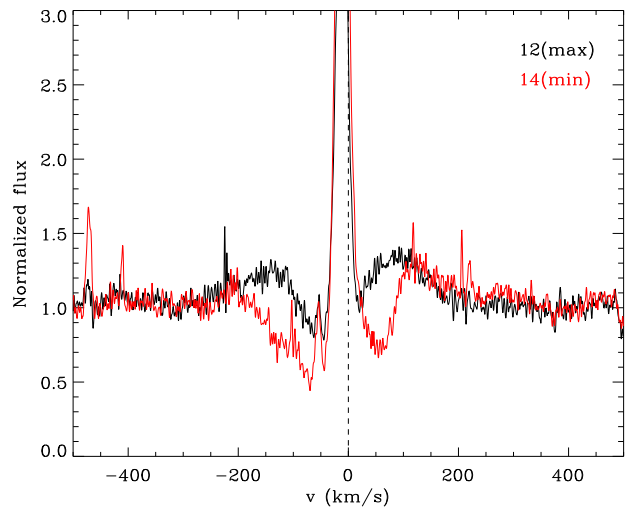}
\caption
{Comparison between \ha circumstellar spectrum observed on March 12 and 14, 
corresponding to photometric maximum and minimum. The dashed vertical line 
indicates the central position of the \ha line. The appearance of a redshifted 
absorption during the brightness minimum of the star is evident.}
\label{f:spec12_14}
\end{figure}

Furthermore, if a hot spot were responsible for the observed photometric 
variations, the light-curve maximum would correspond to the moment when the hot 
spot is fully visible to the observer and the minimum would correspond to the 
opposite situation, when it is completely hidden. The material in free fall in 
the accretion funnel absorbs photons emitted by the hot spot. Thus, the \ha 
redshifted absorption should occur at light-curve maximum, since the spot would 
be in our line of sight at this moment. The opposite is observed in V354 Mon 
spectra, however. The \ha redshifted absorption is visible in the March 14 
spectrum, located at a light-curve minimum, but is not present in the March 12 
spectrum, corresponding to a light-curve maximum, as seen in Fig. 
\ref{f:spec12_14}. According to this result, the hot spot should be facing the 
observer at light-curve minima, which is not consistent with a hot spot as the 
main cause of the photometric variation of V354 Mon. Nevertheless, the existence
of a hot spot is confirmed by the \ha redshifted absorption, since it indicates 
that there is material at high velocity that falls onto the star and hits the 
photosphere, which certainly produces a hot spot at the stellar surface. But
this hot spot does not generate a measurable veiling, at least within the error 
of our measurements, or it may be hidden by the accretion column.

\subsection{Occultation by circumstellar material}

For young stars with an accretion disk seen at high inclination with respect to 
the line of sight, disk material may absorb part of the stellar flux. If the 
material is nonuniformly distributed, the light curve is modulated according to
the disk structure. Inhomogeneities can be caused by azimuthal asymmetries, a 
disk with the outer part partially expanded (flared), or the inner part 
distorted (warped), or even by dust in an inhomogeneous disk wind. In these 
cases, the timescale of photometric variation depends on the disk angular 
velocity at the radius where the inhomogeneity is located. Similarly to spots, 
occultation by circumstellar disk makes the star bluer as it becomes brighter. 
But there is no color variation with the brightness modulation if the disk 
material is completely opaque, as observed in AA Tau 
\citep{bouvier99,bouvier03,bouvier07}. The 3D MHD numerical simulations of 
\citet{romanova04} naturally produced deformations in the inner part of the disk
when the magnetic axis did not coincide with the rotation axis, creating regions
where accretion is favored. This prediction has been confirmed by the study of 
AA Tau, in which a dynamic interaction between the stellar magnetic field and 
the inner disk was observed.

The presence of dust around V354 Mon is indicated by observations with the 
InfraRed Array Camera (IRAC) of Spitzer \citep{teixeira08}, used to identify
infrared excess emission from hot circumstellar dust. The 
$\alpha_{\textrm{\footnotesize IRAC}}$ index represents the slope of spectral energy 
distribution between 3.6 $\mu$m and 8 $\mu$m, which is used to classify the 
inner disk structure, following the criteria proposed by \citet{lada06}. V354 
Mon presents an $\alpha_{\textrm{\footnotesize IRAC}} = -1.72$, indicating that this 
star has an optically thick inner disk. The asymmetric shape of minima and 
maxima observed in the light curve of V354 Mon obtained with CoRoT favors the 
idea that an irregular structure obscures the light emitted by the star, 
probably a circumstellar disk with nonuniformly distributed material, where a 
denser region periodically intercepts the star in our line of sight. This 
phenomenon can only occur if the disk is seen at high inclination. We calculated
the star inclination $i$ through the relation

\begin{equation}
\sin i = \frac {P(v \sin i)}{2 \pi R_*},
\end{equation}

\noindent
where $P$ is the rotation period, $v \sin i$ is the projected rotational 
velocity, and $R_*$ is the stellar radius. Using the effective temperature
and luminosity of V354 Mon given by \citet{flaccomio06} and the PMS theoretical 
evolutionary tracks for solar metallicity computed by \citet{siess00}, we 
inferred a radius of 2.39 R\sun, a mass of 1.49 M\sun, and an age of 
2.4$\times$10${^6}$ years. From a cross-correlation analysis of each spectrum 
with an appropriate numerical mask, we derived a mean rotational velocity of 
$v \sin i = 22.4 \pm 1.2$ \kms. Considering the CoRoT photometric period, 
$P = 5.26 \pm 0.50$ days, we obtained 77\degr\,for the inclination $i$ of the 
stellar rotation axis relative to the line of sight. This means that the system 
presents a high inclination, which supports the possibility of star occultation 
by circumstellar material.

The radius, mass, and age of V354 Mon were also estimated from \citet{landin06},
using a PMS model with non-gray atmosphere, solar metallicity, initially without
rotation and applying mixing length theory with $\alpha = 2$ to treat the 
convective energy transport. The result (Table \ref{t:paramV354}) is slightly 
different from that obtained with the \citet{siess00} model, mainly in mass and 
age, but the values are on the same order of magnitude. 

To investigate the possibility of occultation by circumstellar material as the 
main cause of photometric variability, we applied the model originally developed
for AA Tau \citep{bouvier99} to V354 Mon. The obscuring region is identified as 
a warp, a vertical deformation in the inner part of the circumstellar disk 
produced by the interaction with the dipole magnetic field inclined with respect
to the rotation axis. The occultation model generates a synthetic light curve, 
assuming that the height of the inner disk varies with the azimuthal position 
according to

\begin{equation}
h(\phi)=h_{max}\left |\cos \frac{\pi (\phi - \phi_0)}{2 \phi_c} \right|,
\end{equation}

\noindent
where $\phi_0$ is the azimuth of maximum disk height, corresponding to the 
middle of the eclipse in photometric phase, and $\phi_c$ is the warp azimuthal
semi-extension. Therefore, the height in the inner part of the disk decreases 
gently from its highest value $h_{max}$ at $\phi_0$ to zero at 
$\phi_0 \pm \phi_c$.

Assuming that the photometric period corresponds to the stellar rotation period
and the inner disk warp co-rotates with the star, like in AA Tau, the observed 
photometric period of 5.26 $\pm$ 0.50 days would locate the warp at a distance

\begin{equation}
r_c=\left (\frac{P}{2\pi}\right)^{2/3}(GM_*)^{1/3}=6.1\,R_*
\end{equation}

\noindent
from the star, with $M_*$ and $R_*$ obtained from the model of \citet{siess00}. 
At this distance, the magnetic field affects the accretion dynamics and the disk
is truncated. Internally to this radius, the flux of material is directed to the
star following the field lines. The inclination of the magnetic axis with 
respect to line of sight can be quantified from the redshifted absorptions in 
the emission line profiles because they are produced by the accretion funnel 
material that falls at free-fall velocity in the stellar photosphere. In the 
spectrum of V354 Mon obtained on March 14, located at the photometric minimum, 
we observe an \ha redshifted absorption with a maximum velocity, projected on 
the line of sight, of $125 \pm 25$ \kms. The free-fall velocity of the accretion
flow material from a truncation radius of 6.1 $R_*$ is

\begin{equation}
\textrm{v}_{ff}=\left [\frac{2GM_*}{R_*}\left (1-\frac{R_*}{r_c}\right)\right]^{1/2}\simeq 450 \textrm{ \kms},
\end{equation}

\noindent
with $M_*$ and $R_*$ obtained from the model of \citet{siess00}. Assuming that
the accreting material follows the dipole field lines\footnote{Given the 
inferred mass and age of V354 Mon, it has already developed a small radiative 
core. According to \citet{gregory12}, stars with this configuration present 
large-scale magnetic fields that are axisymmetric and have high-order 
components, typically the octupole, that dominate over the dipole at the stellar
surface. However, the dipolar component is generally stronger than the octupolar
at large distances from the star and then matter flows from the disk towards the
star along dipolar lines. The octupole is more important near the stellar 
surface, redirecting the accretion streams towards the octupolar pole and 
shaping the hot spots 
\cite[see the case of V2129 Oph:][]{donati11,romanova11,alencar12}.}, the 
inclination $i_m$ of the magnetic axis is calculated through the relation 
$\cos i_m = $ v$_{max}/$v$_{ff}$, which gives a value of $\sim 0.28$,
i.e., $i_m \simeq $74\degr. As the inclination of the rotation axis is $\sim$ 
77\degr, the angle between this and the magnetic axis is only a few degrees at 
most. The \citet{romanova03} MHD simulations have shown that for very small 
angles of misalignment between the rotation and the magnetic field axis the 
accretion flows become asymmetric.

We recalculated the inclinations of the rotation and the magnetic field axis 
using the mass and radius values obtained from \citet{landin06} to show the 
dependence of the results with the adopted theoretical evolutionary model. Even 
though the radius values given by the two models are very close, the system 
inclination is affected by the small difference between them. We obtained
$i= 82^\circ$ for $R_*=2.35$ \rs. The inclination of the magnetic field axis with
respect to the line of sight is  $i_m= 71$\degr\,and the misalignment with the 
rotation axis is 11\degr. We note that the differences between the system 
parameters obtained from the two theoretical models (Table \ref{t:paramV354}) 
are negligible, given the expected uncertainties. The results are on the same 
order of magnitude, confirming that the system is seen at high inclination and 
the magnetic axis is only slightly tilted with respect to the rotation axis. The
misalignment between the magnetic and rotation axis is also observed in other 
CTTSs \citep{donati10,donati11}.

\begin{table}[htp]
\caption{V354 Mon parameters obtained from two PMS evolutionary models,
using the effective temperature and luminosity given by \citet{flaccomio06}.}
\label{t:paramV354}
\centering
\begin{tabular}{l c c}
\hline
\hline
Model & \citet{siess00} & \citet{landin06} \\\hline
$R_*$ (\rs)  & 2.39 & 2.35 \\
$M_*$ (\ms)  & 1.49 & 1.12 \\
Age (years)  & $2.4 \times 10^6 $ & $1.2 \times 10^6 $ \\
$i$ & $77^{\circ \: +13^\circ}_{\;\;\, -22^\circ}$ & $82^{\circ \: +8^\circ}_{\: -25^\circ}$ \\
$r_c$ ($R_*$) & $6.1 \pm 1.0 $ & $5.6 \pm 1.1$ \\
v$_{ff}$ (\kms) & $450 \pm 70$ & $390 \pm 80$ \\
$i_m$       & $74^\circ \pm 7^\circ$ & $71^\circ \pm 10^\circ$ \\\hline
\end{tabular}
\tablefoot{We consider the difference between both models as the error in the 
radius and mass determination. Therefore, we calculated the errors of $i$, 
$r_c$, v$_{ff}$, and $i_m$ taking into account the errors in the period, 
rotational velocity, mass, radius, and \ha redshifted absorption v$_{max}$.}
\end{table}

\begin{figure}
\centering
\includegraphics[width=11cm,angle=90]{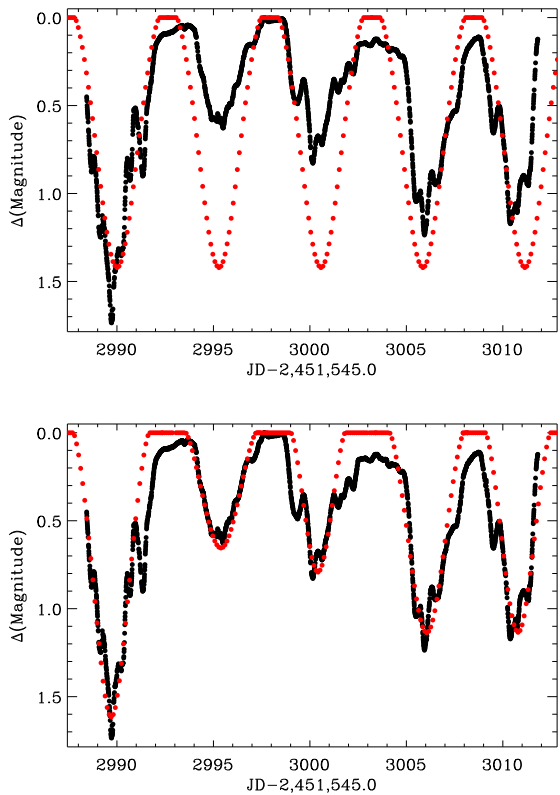}
\caption{Best fit of the occultation model with fixed parameters (top), 
corresponding to a warp with maximum scale height of 0.3 and azimuthal extension
of 360\degr. Individual fit of the model to light-curve minima (bottom). The 
corresponding properties of the warp are indicated in Table 
\ref{t:paramoccm_ind}.}
\label{f:occmodel}
\end{figure}

To reproduce the V354 Mon photometric variability observed with CoRoT 
by employing the occultation model, we considered the mass and radius values 
derived from \citet{siess00}. The model free parameters are the warp maximum 
height $h_{max}$ and azimuthal semi-extension $\phi_c$, which influence the 
amplitude of variability and the duration of the eclipse, respectively. With 
these parameters fixed, the best fit was obtained with $h_{max} =$ 0.3 $r_c$ and 
a total azimuthal extension of 360\degr, represented in Fig. \ref{f:occmodel} 
(top). The synthetic light curve closely follows the large-scale photometric 
variation and reproduces the largest amplitudes observed, but does not fit the 
change in the maxima and minima phase to phase. This is to be expected with 
fixed warp parameters. However, it is interesting to note that the warp 
parameters are similar to those obtained in the model of the variability of
AA Tau, with a maximum height larger than the value traditionally used in disk 
models, $\sim$ 0.05 - 0.1 $r_c$ \citep{bertout88,duchene10}.

The analysis of AA Tau photometric and spectroscopic variability indicated that
the large-scale stellar magnetosphere configuration changed over a month because
of the differential rotation between the star and the disk, showing that the 
circumstellar disk dynamically interacts with the misaligned magnetic field.
Therefore, the deformation in the inner disk, which results from this 
interaction, is also expected to change its characteristics on a short 
timescale. Based on this and the fact that the depth and width of V354 Mon 
photometric minima vary considerably, we modeled each light-curve minimum 
individually. The warp properties are presented in Table \ref{t:paramoccm_ind} 
for each cycle and the corresponding synthetic light curve is shown in 
Fig. \ref{f:occmodel} (bottom). Despite the irregularity of the photometric 
modulation, the parameters of the inner disk warp for each cycle are quite 
similar, indicating that small changes in its height and azimuthal extension can
reproduce the large amplitude of variability in the observed light curve.

\begin{table}[htp]
\caption{Occultation model parameters from individual fit of light-curve 
minima.}
\label{t:paramoccm_ind}
\centering
\begin{tabular}{c c c}
\hline
\hline
Minimum & $h_{max}$ ($r_c$) &  $2 \phi_c$ (\degr) \\\hline
$1^{st}$ & 0.31 & 320   \\
$2^{nd}$ & 0.23 & 320   \\
$3^{rd}$ & 0.25 & 240   \\
$4^{th}$ & 0.28 & 320   \\
$5^{th}$ & 0.28 & 280   \\\hline
\end{tabular}
\end{table}

We derived new values of effective temperature and luminosity using the recently
published spectral type vs. effective temperature/intrinsic color scale of 
\citet{pecaut13}, which is more appropriate for young stars than the 
dwarf-temperature scales used by \citet{flaccomio06}. As the {\it J}-band 
photometry is less affected by the disk and accretion column emissions than 
other bands, we used the available 2MASS {\it J} magnitude \citep{skrutskie06} 
and bolometric correction in this band to compute a luminosity of 
log(L$_{\textrm{bol}}$/\ls) = 0.11. Considering this value and an effective 
temperature of 4330 K, we recalculated the stellar parameters from the PMS 
models of \citet{siess00} and \citet{landin06} (Table \ref{t:paramV354_2}). 
Although the mass and radius have decreased compared with the values derived 
using the effective temperature and luminosity given by \citet{flaccomio06} 
(Table \ref{t:paramV354}), the overall characteristics of the system are on the 
same order of magnitude. This corroborates the scenario of a young star 
surrounded by a disk seen at high inclination, with a small misalignement 
between the magnetic and rotation axis.

\begin{table}[htp]
\caption{V354 Mon parameters obtained from two PMS evolutionary models,
using the effective temperature and luminosity derived from \citet{pecaut13}.}
\label{t:paramV354_2}
\centering
\begin{tabular}{l c c}
\hline
\hline
Model & \citet{siess00} & \citet{landin06} \\\hline
$R_*$ (\rs) & 1.98 & 2.03 \\
$M_*$ (\ms) & 1.09 & 0.88 \\
Age (years) & $2.5 \times 10^6$ & $1.4 \times 10^6$ \\
$i$         & $85^\circ \pm 5^\circ$ & $82^\circ \pm 8^\circ$ \\
$r_c$ ($R_*$) & $6.6 \pm 1.0$ & $6.0 \pm 1.0$ \\
v$_{ff}$ (\kms) & $420 \pm 50$ & $370 \pm 60$ \\
$i_m$        & $73^\circ \pm 7^\circ$ & $70^\circ \pm 9^\circ$ \\\hline
\end{tabular}
\tablefoot{The process to calculate the errors was the same as described in 
Table \ref{t:paramV354}.}
\end{table}

The change in the stellar parameters affected the result of the occultation 
model as the inclination of the system and the corotation radius increased. 
Considering the mass and radius obtained from \citet{siess00}, we reproduced 
exactly the same fit as shown in Fig. \ref{f:occmodel} (top) with fixed 
parameters with a warp of maximum scale height of 0.15 $r_c$ and total azimuthal
extension of 280\degr. The individual fit of the model to light-curve minima 
(Fig. \ref{f:occmodel}, bottom) is also recovered with a warp that changes its 
characteristics from 0.10 to 0.16 $r_c$ in maximum height and from 180\degr\,to
260\degr\,in azimuthal extension. As the inclination of the system is slightly 
steeper in this case, the warp does not need to be too high and too extended to 
generate the observed amplitude of variations. While the stellar parameters 
derived using the scale of \citet{pecaut13} are expected to be more suitable for
this system, the results in the literature are largely based on the relations 
compiled by \citet{kenyon95}. Therefore we kept the results obtained based on 
the latter because they are easier to compare with similar published works.

According to the results obtained, occultation by circumstellar material can be
the main cause of the photometric variability observed in V354 Mon. 
Spectroscopic evidence also favors this interpretation. Photons emitted by the
accretion shock or lower accretion column are absorbed by the funnel material in
free fall, producing a redshifted absorption in H$\alpha$. As seen from Fig. 
\ref{f:spec12_14}, this absorption is more pronounced in the spectrum that 
corresponds to a photometric minimum, but it does not appear in the spectrum 
obtained in the maximum. This is evidence for a correlation between the position
of the accretion flow onto the star and the decreasing of stellar brightness. 
The MHD simulations of \citet{romanova03} indicated that for a misalignment 
between the magnetic field and the rotation axis smaller than 30\degr, the 
densest regions of accreting material to the star are located in two main 
funnels, following the field lines to the closest magnetic pole. Each region is 
located in one hemisphere, one above the disk and the other below, and rotates 
with the star. The region located above the disk periodically occults the star 
for an observer that views the system at medium to high inclination. Therefore, 
the spectroscopic observations and the CoRoT photometry, along with predictions 
of numerical simulations, corroborate the idea of stellar occultation by 
circumstellar material in this system.

A possible explanation for the color variation observed in V354 Mon \BVRI 
photometry is extinction by circumstellar dust. The star becomes redder when 
fainter because the warp may not be totally opaque, different from what was 
observed in AA Tau. An additional analysis is necessary to confirm this
hypothesis as the real cause of V354 Mon color variation, which cannot be done 
based on the data presented in this study.

\section{Conclusions}

From simultaneous high-resolution spectroscopic and photometric observations, 
we analyzed the CTTS V354 Mon. This star exhibits a large, periodic brightness 
variation with minima that change in shape from one rotational cycle to the 
next. A periodogram analysis of the light curve obtained with CoRoT provided a 
photometric period of 5.26 $\pm$ 0.50 days, close to the value derived by 
\citet{lamm05}, 5.22 $\pm$ 0.87 days, which indicates that the main structure 
that produces the photometric modulation did not significantly change over a few
years. Observations at \BVRI bands showed that there is also a small color 
variation and the system becomes slightly bluer as the flux increases.

The spectrum of V354 Mon is variable on a timescale of a few days. The
periodicity of \ha redshifted and blueshifted sides supports the magnetospheric 
accretion scenario, in which the star accretes material from the circumstellar 
disk while it ejects mass through a disk wind that originates close to the 
accretion region.

We investigated the possibility that spots at the stellar surface are the main 
source of photometric variations. A spot with a temperature of $10\,000$ K and 
occupying 5\% of stellar hemisphere is the spot configuration that best 
reproduces the variability amplitudes observed in \BVRI bands, which means that
the possibility of cold spots is discarded. Such a hot spot would produce a 
significant veiling in photospheric lines, which is not observed. In addition, 
the occurrence of pronounced \ha redshifted absorptions seen only in light-curve
minima indicates that the accretion funnel and, consequently, the hot spot are 
visible at this phase, which invalidates this phenomenon as the main cause of 
photometric modulation.

We found evidence that the emission lines vary in a cyclic manner according to 
the photometric modulation. The asymmetric shape of light-curve minima and the 
difference in the spectral line profile at brightness increase and decrease show
that an irregular structure produces these variations, probably a circumstellar 
disk with nonuniformly distributed material. This is supported by the fact that 
the system is seen at high inclination. According to MHD simulations 
\citep{romanova04}, the small misalignment between magnetic and rotational axes 
observed in this star may create a distortion in the inner disk, producing a 
warp that periodically occults part of the stellar photosphere. We used an 
occultation model to determine the general parameters of this structure, which 
exhibits a maximum scale height of 0.3 located near the disk corotation point, 
with an azimuthal extension of 360\degr. These features are similar to those 
obtained in the fit of this model to the variability of AA Tau 
\citep{bouvier99}. The warp in the disk of V354 Mon seems to modify its shape at
each cycle, revealing a dynamical interaction between the stellar magnetosphere 
and the disk inner part, as predicted by MHD simulations 
\citep{goodson99,romanova02}. Nevertheless, the parameters obtained from the 
individual fitting of the model to the light-curve minima are not very 
different from each other. This result shows that, despite the irregularity of 
the brightness modulation, small variations in the warp characteristics are 
capable of reproducing the large amplitude of the observed photometric 
variability. The presence of an \ha redshifted absorption component more 
pronounced only at photometric minimum points to a connection between the 
accretion funnel and the flux variations of the system, which corroborates that 
occultation by circumstellar material is the main cause of photometric 
modulation in V354 Mon.

\begin{acknowledgements}
We thank Suzanne Aigrain for pre-processing the CoRoT light curve and the
anonymous referee for many useful comments and suggestions. NNJF acknowledges 
support from CAPES (fellowship process n$_{o}$.~18697-12-7) and CNPq. SHPA 
acknowledges support from CAPES, CNPq, and Fapemig. JB acknowledges funding from
CNES and from ANR Toupie 2011 Blanc SIMI5-6 020 01. This research was developed 
within the scope of CAPES-Cofecub project and is based on data collected in the 
CoRoT satellite.
\end{acknowledgements}

\bibliographystyle{aa}

\end{document}